\documentclass[conference]{IEEEtran}
\IEEEoverridecommandlockouts
\usepackage{cite}
\usepackage{amsmath,amssymb,amsfonts}
\usepackage{algorithmic}
\usepackage{graphicx}
\usepackage{textcomp}
\usepackage{xcolor}
\usepackage{hyperref}
\usepackage{array, makecell}
\def\BibTeX{{\rm B\kern-.05em{\sc i\kern-.025em b}\kern-.08em
    T\kern-.1667em\lower.7ex\hbox{E}\kern-.125emX}}
\begin{document}
\title{Enhancing MRI-Based Classification of Alzheimer's Disease with Explainable 3D Hybrid Compact Convolutional Transformers\\
}

\author{\IEEEauthorblockN{Arindam Majee*, Avisek Gupta, Sourav Raha, Swagatam Das
\thanks{*Arindam Majee is the corresponding author.}}
\IEEEauthorblockA{\textit{Institute of Advancing Intelligence, TCG CREST}, Kolkata, India \\
\{arindam.majee, avisek.gupta, sourav.raha, swagatam.das\}@tcgcrest.org}
}
\maketitle

\begin{abstract}
Alzheimer's disease (AD), characterized by progressive cognitive decline and memory loss, presents a formidable global health challenge, underscoring the critical importance of early and precise diagnosis for timely interventions and enhanced patient outcomes. While MRI scans provide valuable insights into brain structures, traditional analysis methods often struggle to discern intricate 3D patterns crucial for AD identification. Addressing this challenge, we introduce an alternative end-to-end deep learning model, the 3D Hybrid Compact Convolutional Transformers 3D (HCCT). By synergistically combining convolutional neural networks (CNNs) and vision transformers (ViTs), the 3D HCCT adeptly captures both local features and long-range relationships within 3D MRI scans. Extensive evaluations on prominent AD benchmark dataset, ADNI, demonstrate the 3D HCCT's superior performance, surpassing state of the art CNN and transformer based methods in classification accuracy. Its robust generalization capability and interpretability marks a significant stride in AD classification from 3D MRI scans, promising more accurate and reliable diagnoses for improved patient care and superior clinical outcomes.
\end{abstract}


\section{Introduction}
Alzheimer's disease (AD), a progressive neurodegenerative disorder marked by inexorable cognitive decline and memory loss, presents a burgeoning global public health crisis. According to the United Nations, approximately 55 million individuals worldwide currently live with this devastating illness, with an estimated 6.7 million cases in the United States alone. These staggering figures, projected to escalate to 93 million by 2030 \cite{alzfacts}, emphasize the urgency of addressing this burgeoning epidemic. AD represents the most prevalent form of dementia, accounting for 60-80\% of all dementia cases. This insidious disease unfolds gradually, stealthily infiltrating the mind over years. Early manifestations often involve subtle memory lapses, cognitive difficulties, and impaired learning abilities. As AD progresses, its tentacles reach further, progressively impacting broader areas of the brain. Human memory, thought processes, judgment, language, problem-solving, personality, and even movement can all succumb to its gradual yet relentless advance. Researchers and pathologists demarcate the transitional phase between healthy cognition and AD as Mild Cognitive Impairment (MCI). Currently, no readily available treatment can definitively cure AD or dramatically reverse its pathological effects. However, therapeutic interventions, along with appropriate medication and dedicated care, can potentially mitigate disease progression and delay the onset of debilitating symptoms. Therefore, early diagnosis and comprehensive disease management constitute the cornerstone of mitigating AD's devastating impact.

The timely and accurate diagnosis of AD is paramount for implementing effective intervention and management strategies, minimizing the disease's debilitating impact. Several diagnostic tools play crucial roles in this critical process. One common approach relies on a thorough clinical evaluation by a physician, who meticulously gathers behavioral information. This includes observing cognitive decline, changes in memory, and impairments in daily functioning. The Mini-Mental State Examination\cite{mmse} serves as a valuable tool, providing a standardized assessment of cognitive skills, problem-solving abilities, language comprehension, and motor function through a series of questions and tasks. Based on the patient's performance, the physician assigns a score (out of 30), helping to quantify cognitive decline and monitor disease progression. Beyond clinical evaluation, laboratory tests such as cerebrospinal fluid (CSF) analysis offer additional insights. Measuring the ratio of amyloid and tau proteins in the CSF can provide strong indicators of AD pathology\cite{betatau}, further bolstering the diagnostic picture. Neuroimaging techniques, such as magnetic resonance imaging (MRI) and positron emission tomography (PET), provide invaluable visual representation of brain structure and function. While differentiating between normal age-related changes and AD-related neurodegeneration can be challenging due to potential overlap, these scans offer crucial information for diagnosis, particularly in conjunction with other clinical and laboratory findings. Early and accurate diagnosis of AD remains a complex but critical endeavor. By employing a multifaceted approach that leverages clinical evaluations, CSF analysis, and neuroimaging techniques, clinicians can gain a comprehensive understanding of the disease and personalize management strategies to improve patient outcomes and quality of life. The rapid increase of Alzheimer's cases, coupled with the scarcity of medical resources and the often-subtle early symptoms, demands automated diagnosis tools. Traditional MRI analysis methods, like visual assessments and density measurements \cite{bottino_volume}, struggle with the complexity of 3D data, hindering early detection. Automated algorithms powered by artificial intelligence (AI) offer promising solutions. By analyzing intricate patterns in MRI scans, these algorithms have the potential to diagnose Alzheimer's earlier and more accurately, thereby improving healthcare efficiency and accessibility, especially in resource-limited areas. This concise version maintains the key points while using commonly understood language and avoiding unnecessary technical jargon. It emphasizes the urgency of automated diagnosis and the potential benefits for both patients and healthcare systems.

In the fight against Alzheimer's, researchers are turning to AI as a powerful ally. Machine learning and deep learning techniques are being employed to automate the analysis of 3D MRI scans, aiming to improve the accuracy of AD classification. At the forefront of this effort are  Convolutional Neural Networks (CNNs), renowned for their ability to extract intricate details from images. In the context of AD diagnosis, CNNs excel at identifying subtle changes within localized brain regions \cite{alzhcnn1, alzhcnn2, alzhcnn3, alzhcnn4}. However, their strength in local features can sometimes come at the expense of capturing the broader picture. The complex interplay of brain regions and long-range connections may remain hidden to these powerful yet focused algorithms. This is where the next wave of AI techniques, like transformers, holds promise. Their ability to analyze relationships and dependencies across the entire 3D image could unlock new insights into AD's intricate patterns, paving the way for even more accurate and reliable diagnosis.

Vision Transformers (ViTs) \cite{vit} represent a paradigm shift in computer vision by employing transformer architectures, originally designed for natural language processing, to effectively model visual data without relying on conventional CNNs. These models, inspired by the powerful transformer architecture\cite{attention}, excel at capturing long-range relationships and global patterns in text data. 
ViTs, with their proficiency in capturing distant connections and contextual information, hold great promise for unraveling the complexities of brain tissue and subtle changes indicative of neurological disorders like Alzheimer's disease. However, the application of ViTs directly to 3D MRI data poses unique challenges, given the intricate, non-linear structure of volumetric brain scans. Bridging this divide between language-oriented transformer models and the intricacies of 3D medical data necessitates innovative adaptations to unlock the full potential of ViTs in medical imaging. While ViTs offer a new avenue for decoding information within brain scans, their effective integration into the intricacies of 3D medical data remains an intriguing puzzle.

In response to these challenges, we present the 3D Hybrid Compact Convolutional Transformers (HCCT), a novel deep learning model that seamlessly combines the strengths of CNNs and ViTs. The 3D HCCT excels in capturing both local features and long-range dependencies within 3D MRI data, offering a promising solution for enhanced Alzheimer's disease classification.

Our study introduces the following key contributions in this context:

\begin{itemize}
\item \textbf{Novel 3D HCCT Architecture:} We propose the 3D HCCT architecture, drawing inspiration from the CCT model \cite{cct}. This innovative approach combines the strengths of 3D CNNs and ViTs through our hybrid pooling method, enhancing Alzheimer's classification.
\item \textbf{End-to-End Fully Deep Learning Pipeline:} We present an end-to-end deep learning pipeline for AD diagnosis, encompassing pre-processing steps like skull stripping and standardization, along with classification and explainability analysis, streamlining the workflow and eliminating the need for manual interventions.
\item \textbf{Superior Performance:} The 3D HCCT outperforms leading CNN and ViT-based models on the 3D ADNI dataset, achieving significantly higher accuracy in single-modality (MRI) classification.
\item \textbf{Enhanced Visualization Capabilities:} Our 3D HCCT demonstrates improved visualization capabilities, providing deeper insights into the model's decision-making process.
\end{itemize}

In essence, our work not only addresses the challenges posed by applying ViTs to 3D medical data but also establishes a state-of-the-art model with superior performance and enhanced interpretability for Alzheimer's disease diagnosis.

\section{Related Works}

Methods for AD classification usually consist of an initial pre-processing step to standardize input MRI images, followed by the classification model. A common pre-processing pipeline includes intensity normalization to ensure the MRI voxels/pixels have a common range of values, along with image registration to a standard space such as the Montreal Neurological Institute Space \cite{2011-mni-mri}, and removal of the background and sometimes the skull from the image. While the open-source toolboxes like FreeSurfer \cite{freesurfer}, FSL \cite{fsl1, fsl2, fsl3}, SPM12 toolbox \cite{spm} has often been used for these approaches, \cite{nppy} recently showed that a deep network can be used as an alternative to the entire MRI pre-processing pipeline.

The major challenge that existing deep AD classification approaches have faced has been to effectively model the relevant information from the generally small-sized and imbalanced datasets of 3D MRI images. \cite{2021-jmi} observed feasible transfer learning of ImageNet-pretrained 2D-CNNs for binary classification between AD and Control Normal (CN), with the best performance being achieved when transferring to 3D-CNNs by duplicating the 2D-CNN kernel weights along the third dimension. Transfer learning was also shown to be effective on 2D MRI slices by \cite{2023-ieee-isbi}. Training 3D-CNNs from scratch was carried out for three-class AD/MCI/CN classification by \cite{2022-natscirep}. Here the network's decisions were visualized by generating gradient-based class saliency maps \cite{2014-iclr-simonyan}. Subsequent methods have mostly employed 2D or 3D Grad-CAMs to visualize the decision regions, for example in \cite{2022-medphys}, where a pretrained ResNet18 was fine-tuned on processed 2D axial middle slice images from the 3D MRIs, followed by Grad-CAM visualizations. The inherent problem of class imbalance in AD classification was addressed by \cite{2022-ieeeaccess-addnet}, where an oversampling approach was employed to balance three minority classes of AD corresponding to different progressions, using a four-layer 2D-CNN. Also, obtaining additional feature information from MRIs has been explored in several ways such as wavelet \cite{2023-ieeetbioeng} or Fourier decompositions \cite{2022-miccai-fourier}, using encoded domain knowledge via multiple classifiers \cite{2023-miccai-domain}, or the unsupervised learning of the image domain of MRIs using 3D-DCGANs in \cite{2023-natscirep-3round}. The possibility of training any network for AD classification without assuming data independence, by adversarial training with awareness of data cluster structures was shown to be feasible in \cite{2023-tpami-armed}. A detailed recent study by \cite{2023-ieee-access-alznet} showed that a fine-tuned modified InceptionV3 network excelled at classifying five stages of AD against CN.

Other than convolution-based architectures, GCN approaches have also been explored such as in \cite{2022-miccai-sparse}, where a $k$-NN graph was constructed based on AAL-90 atlas brain ROIs. The Graph Convolutional Network (GCN), decisions were interpreted in terms of top-ranking feature importance probabilities. For similar reasons, feature-graphs were constructed by \cite{2022-miccai-dualgraph} in addition to subject-graphs in a dual-graph setup, to account for noise in the subject data as well. An alternate graph construction by \cite{2022-miccai-intdiff} first graded MRI patches using several 3D U-Nets, followed by constructing a graph from the graded scores at several brain structural locations.

The general efficacy of deep networks for AD classification naturally led to investigations on possible improvements due to attention mechanisms. \cite{2022-kbs} observed improvements due to the use of channel attentions, and likewise success was noted in \cite{2022-ieeeaccess-early} where channel attention and spatial attention were used with the convolution layers. In the approach of \cite{2021-ieeebhi}, after feature extraction from initial 3D convolution layers, a residual self-attention block was used leading to the classification layer. The use of two 3D ResNets with attention, one trained on MRI while the other on PET images, was explored by \cite{2023-kbs}, with final attention blocks at the ends of both network paths further refining the feature representations across all channels. The top performing model observed in \cite{2023-dsp} contained layer blocks consisting of channel attention after convolution layers, followed by the use of transformers before classification. The success of Vis for computer vision problems motivated investigations on their performances, as in \cite{2022-miccai-brainaware}, where a 3D-ViT encoder operated on augmented MRIs to produce several 3D image patches, which were subsequently flattened and provided to an MLP with multi-headed self-attention. The network was trained under a self-supervised contrastive loss which involved transpose convolution layers to reconstruct the MRIs. 2D-ViTs were explored in \cite{2023-ieee-cisce-auxvit}, where augmented MRIs were divided into patches, and fed to the transformer encoder, followed by an attention-aided layer to produce the classification.

Several limitations are inherent to traditional ViT-based models\cite{vit_survey}. Firstly, their patch-based processing approach, while enabling the handling of large images, risks discarding crucial local spatial information. This translates to individual pixel processing, resulting in a significant increase in parameters and computations compared to CNNs. Consequently, computational costs rise in proportion to the parameter count. Additionally, ViTs employ a learnable class token appended to the input sequences, commonly known as the CLS token. This token is fed into the classifier layer, while other sequential outputs are disregarded. This neglects the potential contributions of other sequential outputs of the transformer layers. Within our work, we address these limitations through our proposed novel 3D HCCT architecture.

\section{Proposed Method}
This section presents a comprehensive end-to-end pipeline designed to streamline the processing of raw MRI data and deliver accurate classification results. The pipeline integrates a series of interconnected modules, each serving a distinct purpose
\subsection{Prepossessing 3D MRI scans} Raw MRI scans captures a wealth of information about brain tissues  but also at the same time, it consists of lots of noise such as parts of skull, bones which are not useful for Alzheimer detection. We have segmented out these unnecessary parts. There are lot of open-source tools like FreeSurfer \cite{freesurfer}, FSL \cite{fsl1, fsl2, fsl3}, SPM12\cite{spm} for skull stripping and pre-processing. But major drawback of these tools is they takes lots of time to process a raw image and not suitable for real time application. Xinzi He et al. (2022) had developed a deep learning based tool for MRI pre-processing called Neural Pre-processing Python (NPPY) \cite{nppy}. This is an open-source end-to-end weakly supervised learning approach for converting raw MRI scans into skull-stripped, intensity normalized brain volumes of uniform sizes in the standard coordinate space. We have processed all our input brain MRI scans using NPPY framework and their available pre-trained weights. After, pre-processing we got each MRI volume as skull-stripped, intensity normalized and of $256\times256\times256$ size. To process the information smoothly with our limitted resources we have reshaped these into smaller dimension of $192\times192\times192$. This enables us to make our computation smoothly.

\subsection{3D HCCT architecture}
This section unveils the cornerstone of our proposed pipeline: the 3D HCCT architecture. Recognizing the complementary strengths of CNNs and ViTs, the 3D HCCT seamlessly integrates these two paradigms to achieve superior AD classification performance from 3D MRI data. The architecture's key components are:

\textbf{1. 3D Convolutional Encoder:} The 3D convolutional encoder extracts local features from the input 3D MRI scan. It consists of a stack of convolutional blocks, each block is made of a 3D convolutional layer followed by a 3D batch-normalization layer, a ReLU activation function and finally 3D max pooling layer. The number of convolutional filters are increased at each layer to capture increasingly abstract features. We have fixed kernel size small to capture local dependencies smoothly. The 3D max pooling layer of each convolutional block efficiently down-samples the input image. We took each down-sampled kernel output as a patch. The input 3D image in $192\times192\times192$ produces total 512 patches each of dimension $6\times6\times6$. Later, each of these patches linearly projected in a 1D space i.e., flattened. Finally we perform patch embedding and add a learnable token i.e., CLS token with it.

\begin{figure*}[h] 
    \centering
    \includegraphics[width=\textwidth]{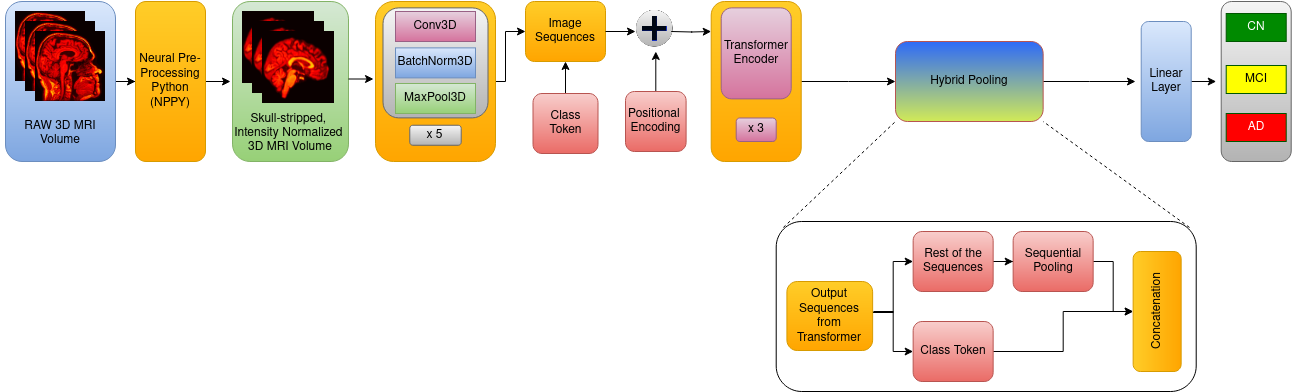}
    \caption{A schematic view of the proposed end-to-end framework}
    \label{fig:proposed-network}
\end{figure*}

\textbf{2. Vision Transformer Encoder:} The ViT encoder is most crucial part of this architecture and plays a vital role to capture both local and long-range dependencies and global interactions within the extracted local features. At its core, the ViT encoder is built upon a series of transformer encoder blocks stacked seamlessly each consisting of a multi-head attention layer and a feed-forward network\cite{attention}. In multi-head attention, the attention head of transformer block is divided into into few smaller attention head each of same size and output of each head is computed parallely. The multi-head attention and self attention layer allows the ViT encoder to learn relationships between different parts of the input feature sequence, while the feed-forward network injects non-linearity into the model. There are Dropout\cite{dropout} layer in between these linear layers to control the over-fitting.

\textbf{3. Hybrid Pooling:} To map the sequential outputs to the output class index, ViT uses a learnable class or query token through the network and later feeds it it to the classifier. Hasani et al. introduced SeqPool\cite{cct}, an attention-based method which pools over the output sequence of tokens. Their motivation for SeqPool was that the transformer encoder's output sequence contains relevant information across different parts of the input image, therefore preserving this information can improve performance. Instead of a learnable CLS token at each transformer block, this SeqPool requires less parameter which essentially reduces computation cost. Taking inspiration from SeqPool we have developed a novel hybrid method to leverage the advantages of sequential pooling along with traditional class tokens used in ViTs. In our proposed method this hybrid operation consists of mapping the output sequence using the transformation $T: \mathbb{R}^{b \times (n+1) \times d} \to \mathbb{R}^{b \times d}$
Given:
\[
x_L = f(x_0) \in \mathbb{R}^{b \times (n+1) \times d},
\]

where $x_L$ is the output of an $L$ layer transformer encoder $f$ , $b$ is batch size, $n$ is number of patches, $(n+1)$ is the total sequence length considering class token, and $d$ is the hidden embedding dimension. The class token, $x_c \in \mathbb{R}^{b \times 1 \times d}$ is separated from this $x_L$ and the rest of the part, $x_a \in \mathbb{R}^{b \times n \times d}$ is fed to a linear layer $g(x_a) \in \mathbb{R}^{d \times 1} $, and softmax activation is applied to the output:
\[
x_{a}^{'} = softmax(g(x_a)^T) \in \mathbb{R}^{b \times 1 \times n}.
\]
This generates an importance weighting for each input token, which is applied as follows:
\[
x_{a}^{''} = x_{a}^{'} \times x_{a}  \in \mathbb{R}^{b \times 1 \times d}.
\]
Now, this is further concatenated with $x_{c}$ as follows
\[
z = concat [ x_{c}, x_{a}^{''} ] \in \mathbb{R}^{b \times 1 \times 2d}.
\]
By flattening, the output $z \in \mathbb{R}^{b \times 2d}$ is produced. This output is now sent through a classifier. In our experiments the classifier is a single linear layer $g_{c}(z) \in \mathbb{R}^{2d \times C}$, where $C$ is the number of output classes.

\section{Experiments}
\subsection{Dataset Description}
To rigorously evaluate the performance of the proposed 3D HCCT architecture for AD classification from 3D MRI scans, we leverage the widely recognized Alzheimer's Disease Neuroimaging Initiative (ADNI) dataset\cite{adni}. This comprehensive dataset provides access to a large collection of MRI scans from individuals diagnosed with AD, MCI, and CN. For our experiments, we curated a dataset from ADNI-I, consisting of a total of 2182 3D MRI scans. This dataset is further stratified into three distinct sets: training, validation, and test. The distribution of scans across these sets, categorized by clinical diagnosis (CN, MCI, AD), is detailed in Table \ref{tab:MRI scan distribution across each set}

\begin{table}[]
    \centering
    \caption{MRI scan distribution across each set}
    \begin{tabular}{|c|c|c|c|c|}
         \hline
         Class &  CN & MCI & AD & Total \\
         \hline
         Train & 523 & 686 & 317 & 1526\\
         \hline
         Validation & 112 & 147 & 67 & 326 \\
         \hline
         Test & 113 & 148 & 69 & 330 \\
         \hline
         Total & 748 & 981 & 453 & 2182 \\
         \hline
    \end{tabular}
    \vspace{4pt}
    \label{tab:MRI scan distribution across each set}
\end{table}

\begin{figure}
    \centering
    \includegraphics[scale=0.43]{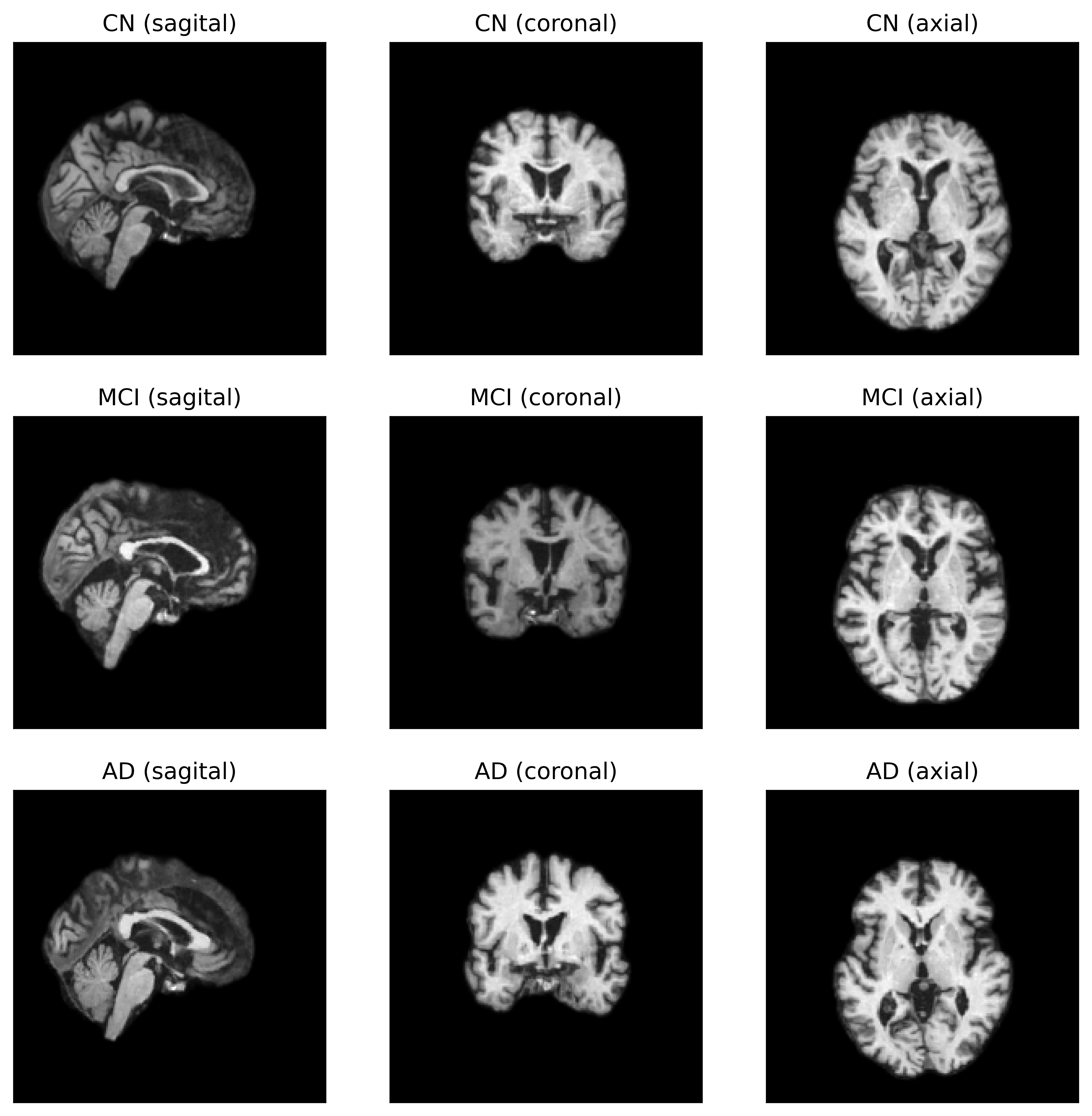}
    \caption{Sagital, coronal and axial view of a sample image from each class of ADNI dataset}
    \label{fig:enter-label3}
\end{figure}

\subsection{Experimental Setup}
The training and evaluation of the proposed pipeline were conducted on a single NVIDIA GeForce RTX-4090 GPU. To expedite the training process, batch-wise training was employed, with each batch containing three MRI volumes. The model was trained for 100 epochs using the AdamW optimizer\cite{AdamW}, resulting in a model referred to as the "Base" model. Further refinement was achieved through fine-tuning. Specifically, the transformer layers were frozen, and only the patch embedding and classifier layers were fine-tuned for an additional 50 epochs with a reduced learning rate. This "Fine-Tuned" model served as the basis for subsequent analyses.

To assess the impact of the network architecture on performance, ablation studies were conducted on the ADNI dataset by varying the number of transformer encoder layers. The best-performing model was subsequently evaluated on additional datasets to assess its generalizability. For detailed information on the experiment setup, hyperparameters, and code, please refer to the provided repository: \href{https://anonymous.4open.science/r/Diagnosis-of-Alzheimer-with-3D-Hybrid-Compact-Convolutional-Transformers-from-MRI-C1D6}{https://anonymous.4open.science/r/Diagnosis-of-Alzheimer-with-3D-Hybrid-Compact-Convolutional-Transformers-from-MRI-C1D6}

\begin{table}[]
    \centering
    \caption{Hyper-parameter Details}
    \begin{tabular}{|c|c|} 
        \hline
        \textbf{Hyper-parameter} & \textbf{Value} \\ \hline
        No. of epochs & 100 \\ \hline
        Dropout & 0.2 \\ \hline
        Learning Rate	& 4e-5 \\ \hline
        Learning Rate Decay Method	& Step decay \\ \hline
    \end{tabular}
    \vspace{4pt}
    \label{tab:table2}
    \vspace{-5pt}
\end{table}


\subsection{Results}
The effectiveness of our proposed 3D HCCT network for AD classification was evaluated using a comprehensive set of metrics, including accuracy, precision, recall, and F1-score. Each metric provides a distinct perspective on the model's performance, allowing for a nuanced understanding of its strengths and limitations. On the ADNI dataset, the 3D HCCT achieved an outstanding classification accuracy, as detailed in Table \ref{tab:Performance of HCCT on ADNI Dataset}.  The table presents the performance of both the "Base" model, trained with the full network architecture, and the "Fine-tuned" model, where only the patch embedding and classifier layers were further fine-tuned.

Further analysis of these results, along with detailed breakdowns of the different metrics for both models, will be presented in the following subsections. This analysis will delve into the specific contributions of each component of the architecture and provide insights into the model's generalization capabilities across diverse datasets. 

\begin{table}[]
    \centering
    \caption{Performance of HCCT on ADNI Dataset}
    \resizebox{\columnwidth}{!}{
    \begin{tabular}{|c|c|c|c|c|c|c|}
        \hline
        Model & \makecell{No. of  \\ Encoder \\ Layers} & \makecell{No. of \\parameters \\ (in M)} & \makecell{Accuracy\\ (\%)} & Precision & Recall & F1-score \\  \hline
        Base & 3 & \textbf{6.22} & \textbf{95.76} & 0.9578  &  0.9576  &  0.9576 \\  \hline
        Fine-Tuned & 3 & \textbf{\textcolor{blue}{6.22}} & \textbf{\textcolor{blue}{96.06}}  & 0.9606  &  0.9606  &  0.9606\\  \hline
        Base & 4 & 6.69 & 92.12 & 0.9246  &  0.9242  &  0.9240\\   \hline
        Fine-Tuned & 4 & 6.69 & 94.84 & 0.9492  &  0.9485  &  0.9484 \\  \hline
        Base & 5 & 7.16 & 94.84 & 0.9547  &  0.9545  &  0.9545 \\  \hline
        Fine-Tuned & 5 & 7.16 & 95.16 & 0.9557  &  0.9545  &  0.9545 \\  \hline
        Base & 6 & 7.63 & 94.54  & 0.9434  &  0.9424  &  0.9423\\  \hline
        Fine-Tuned & 6 & 7.63 & 95.45 & 0.9524  &  0.9515  &  0.9515 \\  \hline
    \end{tabular}}
    \vspace{4pt}
    \label{tab:Performance of HCCT on ADNI Dataset}
\end{table}


\begin{figure}
    \centering
    \includegraphics[scale=0.4]{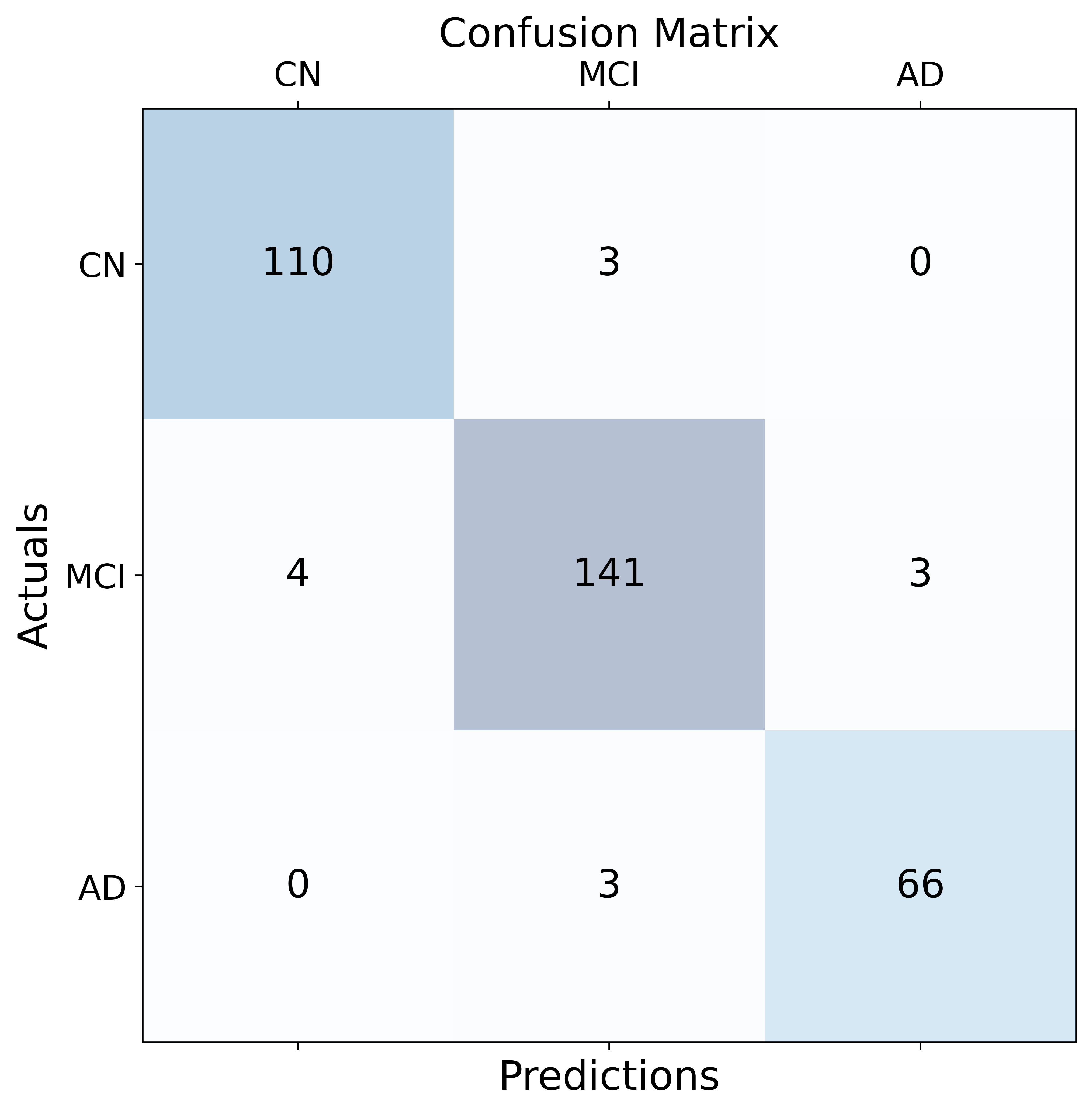}
    \caption{Confusion Matrix of Test set for HCCT Finetuned model with 3-layer transformer encoder}
    \label{fig:confusion_matrix}
\end{figure}

\subsection{Comparison with State-of-the-Art Methods}
To assess the efficacy of the proposed 3D HCCT network, we conducted a comprehensive evaluation against prominent CNN and ViT-based methods for Alzheimer's disease (AD) classification. As shown in Table \ref{tab:comparision}, the 3D HCCT exhibits remarkable performance, exceeding the capabilities of existing approaches in various metrics. Overall, the 3D HCCT's superior performance in comparison to state-of-the-art methods establishes its potential as a powerful tool for AD classification. Its accuracy and efficiency make it a promising candidate for real-world clinical applications, potentially enabling earlier and more accurate diagnoses of AD and paving the way for improved patient outcomes.

\begin{table}[]
    \centering
    \caption{Comparison with the existing methods}    
    \label{tab:comparision}
    \resizebox{\columnwidth}{!}{
    \begin{tabular}{|c|c|c|c|c|}
        \hline
        Methods & Year & No. of classes & Accuracy  \\ \hline
        DGLCN\cite{2022-miccai-dualgraph} & 2022 & 2 & 86.7 \\ \hline
        GF-Net\cite{2022-miccai-fourier} & 2022  & 2 & 94.1 ± 2.8  \\ \hline
        \makecell{Learning with\\ Domain-Knowledge}\cite{2023-miccai-domain} & 2023 & 2 & 86.2 \\ \hline
        Aux-ViT\cite{2023-ieee-cisce-auxvit} & 2023 & 2 & 89.58 \\ \hline
        3D DCGAN\cite{2023-natscirep-3round} & 2023 & 2 & 92.8 \\ \hline
        Multimodal 3D CNN \cite{2023-kbs} & 2023 & 2 & 94.61 \\\hline
        \textbf{HCCT (our)} & \textbf{2024} & \textbf{3} & \textbf{96.06} \\ \hline
    \end{tabular}}
\end{table}

\vspace{-3pt}

\subsection{Visualization of Feature Importance}
To gain deeper insights into how the 3D HCCT extracts crucial features, we visualized the areas of the 3D MRI scan that most heavily influence its decision-making process. We focused on the attention probabilities within the transformer encoders, as these indicate the weight given to each small "patch" of the image. Averaging these probabilities across all transformer layers and individual heads created a score reflecting each patch's importance. Next, we combined these scores with the feature activations, the outputs of the 3D convolutional encoder, averaged across all channels. This generated a "heatmap" highlighting the regions that contribute most to the model's predictions. To visualize this heatmap on the original scan, we upsampled it to match the original image size.

\begin{figure*}[h] 
    \centering
    \includegraphics[width=0.9\textwidth]{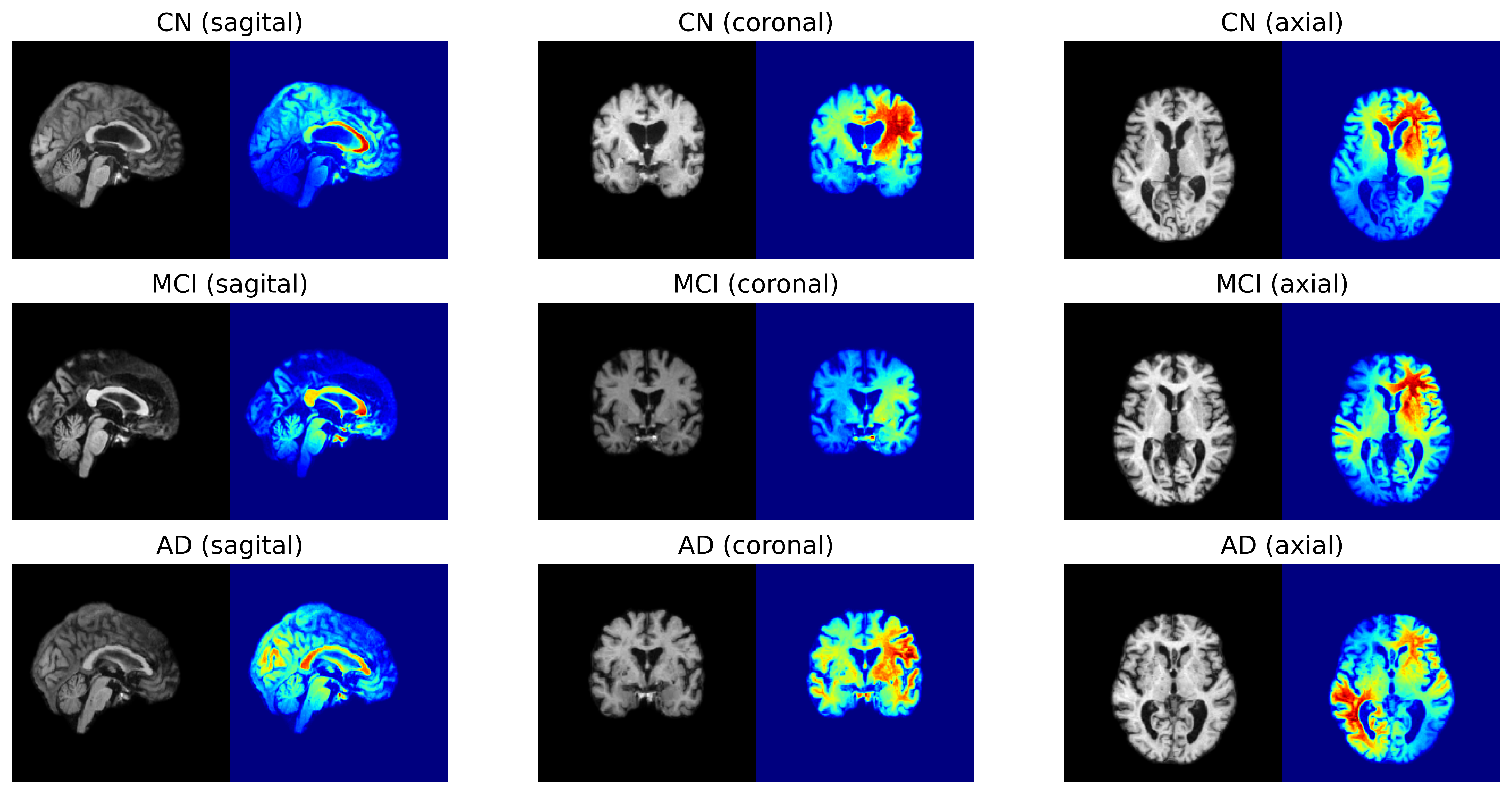}
    \caption{Heat-map visualization of each class}
    \label{fig:visualization}
\end{figure*}

Importantly, in MRIs, pixel intensity reflects brain tissue presence. Therefore, to identify the relevant anatomical areas, we multiplied the original brain scan with the heatmap. This rectified the image and excluded non-tissue regions. Figure\ref{fig:visualization} displays this visualization for one image from each class (CN, MCI \& AD) in sagittal, coronal, and axial views. Hotter (red) regions indicate areas that contribute more significantly to the classification, offering valuable insights into the 3D HCCT's decision-making process.

\subsection{Discussion}
The proposed 3D HCCT architecture shines in its ability to accurately classify Alzheimer's disease (AD) from 3D MRI scans. Our experiments reveal fascinating nuances in its performance as the number of transformer layers changes. While some variation exists, the 3-layer transformer consistently tops its peers. Interestingly, pushing the layer count beyond this point leads to a subtle decline in performance. This hints at potential overfitting, where additional layers introduce more complexity without a compensating boost in accuracy. This trend holds true for both base and fine-tuned models.

A significant strength of our approach lies in its impressive generalizability even with an imbalanced training dataset. This robustness bodes well for its adaptability to real-world scenarios where data distributions might be skewed. Analyzing various metrics in tandem provides valuable insights into the model's strengths and limitations. For instance, a model with high accuracy but low recall might excel at avoiding false positives while potentially overlooking true AD cases. Conversely, a model with high recall but low precision might flag most AD cases but also generate many false alarms. Choosing the most relevant metric, therefore, hinges on the specific context and priorities of the research. In the critical domain of AD classification, where early diagnosis is paramount, a high recall score might trump high accuracy. Even a few missed AD cases could have devastating consequences for patients, making sensitivity a crucial factor in evaluating model performance.

These results illuminate several key advantages of the 3D HCCT. Its ingenious combination of 3D convolution and transformers allows for effective capture of both local and global features within the MRI data. Furthermore, the it' efficient parameterization and hierarchical architecture contribute to superior performance with reduced computational demands. This makes the network well-suited for resource-constrained settings, a valuable asset in real-world healthcare applications.

\section{Conclusion}

In conclusion, the introduction of the 3D Hybrid Compact Convolutional Transformer (3D HCCT) marks an advancement in the landscape of Alzheimer's disease (AD) classification from 3D MRI scans. The strategic fusion of CNNs and ViTs within the 3D HCCT demonstrates a sophisticated capacity to capture both intricate local details and long-range relationships within the MRI data, resulting in a notable performance leap over state-of-the-art CNN-based methods on ADNI datasets. Significantly, the 3D HCCT attains heightened accuracy, sensitivity, and specificity in identifying AD cases, deriving its success from the extraction of richer and more discriminative features, a direct outcome of its dual focus on local and long-range dependencies within the MRI data. Furthermore, the 3D HCCT's commendable generalizability is underscored by its consistent strong performance across diverse datasets, affirming its robustness and adaptability to real-world clinical scenarios. These compelling findings emphasize the considerable potential of the 3D HCCT as a potent tool for AD classification, promising earlier and more reliable diagnoses. This advancement not only holds the prospect of improved patient care but also beckons exciting avenues for future research in deep learning-based AD classification methods.

As we peer into the future, numerous promising paths beckon our exploration. Venturing into the utilization of alternative deep learning architectures, such as the recurrent neural network Graph Convolutional Network (GCN), to capture temporal and spatial relationships within MRI data, and the development and validation of deep learning models effective in real-world clinical settings, stand out as pivotal challenges. Addressing these challenges and harnessing the power of deep learning undoubtedly propel the boundaries of AD classification, ultimately translating to enhanced patient care and improved outcomes for those grappling with this devastating disease.

\section*{Acknowledgment}

We extend our sincere gratitude to several entities who played crucial roles in the successful completion of this work. The Alzheimer's Disease Neuroimaging Initiative (ADNI) deserves our deepest appreciation for providing the neuroimaging data that formed the foundation of our experiments. Furthermore, we are grateful for the invaluable insights and expertise offered by Dr. Young Li from Biomax, Labvantage, and Prof. Sumantra Chatterji from the Centre for High Impact Neuroscience and Translational Applications, TCG CREST. 


\bibliographystyle{ieee} 
\bibliography{ijcnn_alzheimer} 

\end{document}